\documentclass[sigconf]{acmart}

\newcommand*{\fixmeON}{} 

\newcommand{\user}[3]{{%
#2#3-#1}}

\newcounter{boldifyCounter}
\newcounter{fixmeSectionCounter}
\newcounter{fixmeTotalCounter}
\makeatletter
\@addtoreset{fixmeSectionCounter}{section}
\@addtoreset{fixmeSectionCounter}{subsection}
\@addtoreset{boldifyCounter}{section}
\@addtoreset{boldifyCounter}{subsection}
\makeatother

\newcommand{\boldify}[1]{}
\ifdefined\boldifyON
	\renewcommand{\boldify}[1]{\par\noindent%
		\stepcounter{boldifyCounter}%
		\textbf{{\color{green}**}%
		~\arabic{section}.\arabic{subsection}.\arabic{boldifyCounter}%
		: #1}
	}
\fi

\newcommand{\FIXME}[1]{}
\ifdefined\fixmeON
	\renewcommand{\FIXME}[1]{\par\noindent%
		\stepcounter{fixmeSectionCounter}\stepcounter{fixmeTotalCounter}%
		{\color{red}\fbox{\color{black}%
			\parbox{.97\linewidth}{%
				\textbf{FIXME \arabic{section}.\arabic{subsection}.%
        		\arabic{fixmeSectionCounter} (\color{red}%
        		\#\arabic{fixmeTotalCounter}):} #1}}%
        }
	}
\fi

\makeatletter
\@ifclasswith{acmart}{review}{
    \usepackage{fancyhdr, datetime}
    \pagestyle{fancy}
    \fancyfoot{}
    \pagenumbering{arabic}
    \rfoot{\smallbreak{\tiny\today~\currenttime~GMT\quad\quad}Page~\thepage~}
    
}{}
\makeatother

\usepackage{calc, enumitem} 
\usepackage{amssymb, wasysym}

\copyrightyear{2019} 
\acmYear{2019} 
\setcopyright{acmcopyright}
\acmConference[IUI '19]{24th International Conference on Intelligent User Interfaces}{March 17--20, 2019}{Marina del Rey, CA, USA}
\acmBooktitle{24th International Conference on Intelligent User Interfaces (IUI '19), March 17--20, 2019, Marina del Ray, CA, USA}
\acmPrice{15.00}
\acmDOI{10.1145/3301275.3302310}
\acmISBN{978-1-4503-6272-6/19/03}

\begin{document}

\textbf{\LARGE{CHANGELOG}}
\begin{enumerate}
\item \textbf{\color{red}(DONE)} \textbf{Authorship text} -- Including short title (located in 000-titleauthorabstract)

\item \textbf{\color{red}(DONE)} \textbf{WHEN study was run} -- R2 wanted this information added for archival purposes. Added a few words to start of section 5 paragraph 2 (In September 2018)

\item \textbf{\color{red}(DONE)} \textbf{Move RQs} -- R4 suggested placing the RQs at the start of the section as opposed to the end

\item \textbf{\color{red}(DONE)} \textbf{Confusing Quote} -- R1: ``race was not a predictor may not accurately reflect the reality'', but perhaps is supposed to read `` [if] race was not a predictor [it] may not accurately reflect the reality''

\item \textbf{\color{red}(DONE)} \textbf{Acknowledge} -- Added, acknowledging Bhanu and DARPA

\item \textbf{\color{red}(DONE)} \textbf{Participants saw what EXACTLY?} -- R4/1 requested that we clarify this.\\
\textbf{\color{blue}(FIX)} Clarified in caption of Figure 1.  Will also do so in the supp materials.

\item \textbf{\color{red}(DONE)} \textbf{Confusing Sentence} -- R2: on p8 it says that 33.1\% of participants thought using race as a factor was either fair or unfair -- but the way the sentence is written, I could not tell which it was!\\
\textbf{\color{blue}(FIX)} Adjusted sentence a bit to have fewer conjunctions and disjunctions. I dont think it was that unclear...

\item \textbf{\color{red}(DONE)} \textbf{Definitions} -- R1: The authors do not fully define an individual's need for cognition; \\
\textbf{\color{blue}(FIX)} We don't actually use this much, so it seems ok to punt the definition to an external document, as cited at the end of section 3.

\item \textbf{\color{red}(DONE)} \textbf{what group was p107 in?} -- R1 wants to know what treatment info in order to contextualize quotes. \\
\textbf{\color{blue}(FIX)} I decided it was better to bake this into ALL quotes (ie. RI-P107) and an accompanying footnote explaining the syntax. 
This is done via command, so it is easy to undo or reformat however. 
All changes are localized to section 7

\item \textbf{\color{red}(DONE)} \textbf{Tie background to RQs} -- R4 request.\\
\textbf{\color{blue}(FIX)} Chopped the last sentence off of each background subsection and added another sentence to clarify the gap our research fills.
Also added a sentence above the RQs pointing back to those identified gaps.

\item \textbf{\color{red}(DONE)} \textbf{Tie overview to RQs} -- R4 request.\\
\textbf{\color{blue}(FIX)} I think this was dealt with by clarifying \textbf{Tie background to RQs}

\item \textbf{\color{red}(DONE)} \textbf{Why THESE explanation styles} -- R4/R1 wanted this information added to study overview section where we discuss Binns et al's work. 
In addressing this, consider discussing why we didnt expose confidence info (confound we could not control). 
Also consider discussing why we showed outputs 1 at a time (ecological validity???)\\
\textbf{\color{blue}(FIX)} Reworked Explanation Styles paragraph of the study overview to clarify the justifications.
The confidence part comes up later in the Limitations section of Discussion.

\item \textbf{\color{red}(DONE)} \textbf{Taxonomy} -- R1: utilize prior terminology in their explanation discussion. For example, the sensitivity based explanation is most similar to a `what if,' and the case-based explanation is the least `sound'.\\
\textbf{\color{blue}(FIX)} Rolled in with \textbf{Why THESE explanation styles}

\item \textbf{\color{red}(DONE)} \textbf{Threats to validity} -- R4/2 requested a short bit added to the discussion section on threats sourced from the MTurk sampling\\
\textbf{\color{blue}(FIX)} Added subsection to the end of discussion

\item \textbf{\color{red}(DONE)} \textbf{Further Discuss case-based (rel. to Taxonomy)} -- R1 wants us to discuss how case-based explanations have lower soundness than the others, partially due to ``explanation of the explanation''\\
\textbf{\color{blue}(FIX)} Added to the footnote about insufficient justification in section 7

\item \textbf{\color{red}(DONE)} \textbf{Tie Binns to results} -- R1's reasonable request. JED would summarize Binns results in the following bullet points:
\begin{enumerate}
    \item lack of human touch/negotiation 
    \\\textbf{\color{blue}(FIX)} corroborated briefly by section 7 ``general trust or distrust in ML systems)
    \item (un)acceptability of statistical process 
    \\\textbf{\color{blue}(FIX)} corroborated by section 7 ``general trust or distrust in ML systems)
    \item Actionability of explanations is important 
    \\(not observed). Likely due to difference in use case perspective. JED votes we say nothing about this.
    \item Some features unaccounted for
    \\ \textbf{\color{blue}(FIX)} corroborated by section 7 ``Lacking features)
    \item Loose definitions of moral constructs
    \\ \textbf{\color{blue}(FIX)} corroborated by the discussion surrounding Figure 3. Fix is applied in section 8 ``Individual differences and descriptive fairness'')
    \item Between subjects study found few significant differences between explanation treatments
    \\ (not observed, but I think we have beaten the dead horse about our superior experiment design sufficiently to edit nothing in response to this)
    \item Within subjects study found that case-based explanation performs poorly
    \\ \textbf{\color{blue}(FIX)} corroborated by section 7 ``Case-based)
    
\item \textbf{\color{red}(DONE)} \textbf{Figure Label} -- R1: the top label on Figure 4 is strange (I think because it is a double negative ``false for low ML fairness'' to mean ``trusts ML''), perhaps it should instead be ``Trust in ML'' ``yes'' ``no''
\\ \textbf{\color{blue}(FIX)} Figures have been reformatted wholesale.
Legends have been removed, sizes adjusted, etc.
Output looks visually similar to what was shown in the paper
\end{enumerate}

\item \textbf{\color{red}(DONE)} \textbf{Add descriptions} (new template requires it)

\item \textbf{\color{red}(DONE)} \textbf{Article Audit} -- R2 noted some minor typos around missing articles. JED will be doing copy editing tomorrow, can check at that time

\item \textbf{\color{red}(DONE, but badly)} \textbf{CCS codes} JED added the most obvious one, if anyone wants to improve upon it, feel free

\item \textbf{\color{red}(DONE, but badly)} \textbf{keywords} JED took a stab at these, feel free to improve
\end{enumerate}

\textbf{\LARGE{TODO-LIST}}
\begin{enumerate}

\item \textbf{Check CCS codes and keywords} JED took a stab at these, please examine

\item \textbf{Finalize copyright info} JED currently assumes no one wants to pay the open access fee and that ACM managed copyright is preferable. Need confirmation before finalizing this piece

\item \textbf{Typography} -- Once Keywords/copyright are dealt with JED can start making the pages look nice

\item \textbf{How to post (or repost) supp materials} -- JED and Vera discussed this. 
We can upload with the manuscript and it will be hosed on ACM Digital Library, but there are advantages to duplicate hosting (I do not have good permanent hosting).\\
\textbf{\color{purple}(VL TODO)} I could not find a good solution for hosting either. Lets just go with the default ACM solution\\ 
\textbf{\color{purple}(JED)} Affirmative. Leaving this item on the TODO list to ensure I remember to upload the supplement\\ 
\end{enumerate}

\clearpage

\setcounter{page}{1}

\title{Explaining Models: An Empirical Study of 
How Explanations Impact Fairness Judgment}

\author{Jonathan Dodge}
\email{dodgej@eecs.oregonstate.edu}
\affiliation{Oregon State University}

\author{Q. Vera Liao}
\email{vera.liao@ibm.com}

\author{Yunfeng Zhang}
\email{zhangyun@us.ibm.com}
\affiliation{IBM Research AI}

\author{Rachel K. E. Bellamy}
\email{rachel@us.ibm.com}

\author{Casey Dugan}
\email{cadugan@us.ibm.com}
\affiliation{IBM Research AI}

\renewcommand{\shortauthors}{J Dodge et al.}

\renewcommand{\shorttitle}{An Empirical Study of 
How Explanations Impact Fairness Judgment}

\begin{abstract}
Ensuring fairness of machine learning systems is a human-in-the-loop process.
It relies on developers, users, and the general public to identify fairness problems and make improvements.
To facilitate the process we need effective, unbiased, and user-friendly explanations that people can confidently rely on. 
Towards that end, we conducted an empirical study with four types of programmatically generated explanations to understand how they impact people's fairness judgments of ML systems.
With an experiment involving more than 160 Mechanical Turk workers, we show that:
1) Certain explanations are considered inherently less fair, while others can enhance people's confidence in the fairness of the algorithm; 2) Different fairness problems--such as model-wide fairness issues versus case-specific fairness discrepancies--may be more effectively exposed through different styles of explanation; 3) Individual differences, including prior positions and judgment criteria of algorithmic fairness, impact how people react to different styles of explanation. We conclude with a discussion on providing personalized and adaptive explanations to support fairness judgments of ML systems.
\end{abstract}


%

\maketitle

\section{Introduction}

Increasingly, important decisions that impact human lives and societal progress are supported by machine learning (ML) systems. Examples where ML systems are used to make decisions include  hiring, marketing, medical diagnosis, and criminal justice. 
This trend gives rise to concerns about algorithm fairness---or possible discriminatory consequences for certain groups of individuals. 
Machine learning algorithms are trained based on data from past decisions, decisions which may have themselves been biased and discriminatory. 
Research shows that by optimizing for the unitary goal of accuracy, ML algorithms trained on historical data not only replicate, but may amplify existing biases or discrimination~\cite{zhao2017men}. 
The possibility of spiraling discriminatory consequences is driving a distrust and ``fear of AI'' in public discussions (e.g.,~\cite{RacistAI,WhiteAI}).

There is a growing body of work on developing non-discriminatory ML algorithms (e.g.,~\cite{kamishima2012fairness,joseph2016fairness,zafar2017fairness}), equal attention has not been paid to the human scrutiny necessary to identify and remedy fairness issues.
The need for such research is highlighted by recent studies, which show that algorithmic fairness often may not be prescriptively defined, but is multi-dimensional and context-dependent~\cite{Grgic-Hlaca2018}. 
Public scrutiny of the usage of risk assessment algorithms in the criminal justice system~\cite{ProPublicaData,Larson2016} brings attention to the need to progress the accountability and fairness of such algorithms.

Accurately identifying fairness issues in ML systems is extremely challenging, however. 
Most ML algorithms aim to produce only prediction or decision outcomes, while humans tend to rely on information about decision-making \textit{processes} to justify the decisions made.
ML algorithms are often seen as ``black boxes'', where one can only see the output and make a best guess about the underlying mechanisms.
This problem is further exacerbated by the popularity of deep learning algorithms, which are often unintelligible even for experts.
This lack of transparency drives a sweeping call for \textit{explainable artificial intelligence} (XAI) in industry, academia, and public regulation. 
For example, the EU General Data Protection Regulation (GDPR) requires organizations deploying ML systems to provide affected individuals with meaningful information about the logic behind their outputs.

Critically, explanations are not just for people to understand the ML system, they also provide a more effective interface for the human in-the-loop, enabling people to identify and address fairness and other issues. 
When people trust the explanation, it follows that they would be more likely to trust the underlying ML systems.

Much research is dedicated to generating explanations in various styles, including model-agnostic approaches~\cite{ribeiro2016should,lundberg2017unified} applicable to any ML algorithm.
However, this body of research is criticized for ``approaching this [XAI] challenge in a vacuum considering only the computational problems''~\cite{miller2017explanation} without the quintessential understanding of how \emph{people} perceive and use the explanations.

In this paper, we conduct an empirical study on how people make fairness judgments of ML systems and how explanation impacts that judgment.
We aim to highlight the nuances of such judgments, where there are different types of fairness issues, different styles of explanation, and individual differences, to encourage future research to take more user-centric and personalized approaches.

Specifically, we identify four styles of explanation based on prior XAI work and automatically generate them for a ML model trained on a real-world data set.
In the experiment, we explore the effectiveness of explanations in exposing two types of fairness issues--model-wide unfairness produced by biased data, and fairness discrepancies in cases from different regions of the feature space.
Our user study demonstrates that judging fairness is not only influenced by explanation design, but also an individual's prior position on algorithmic fairness, including both the general trust of ML systems for decision support and one's position on using a particular feature.
We also present user feedback for the four styles of explanation.
Our results provide insights on the mechanisms of people's fairness judgment of ML systems, and design guidelines for explanations to facilitate fairness judgment making.
We first review relevant work, then present the study overview and research questions.

\section{Background}

\subsection{Fairness of Machine Learning Systems}
One of several definitions for algorithmic fairness is: ``\textit{...discrimination is considered to be present if for two individuals that have the same characteristic relevant to the decision making and differ only in the sensitive attribute (e.g., gender/race) a model results in different decisions}''~\cite{Calders2013}.
The consequence of deploying unfair ML systems could be \textit{disparate impact},  practices which adversely affect people of one protected characteristic more than another in a comparable situation~\cite{Calders2013, Grgic-Hlaca2018}.

Despite the ``statistical rationality'' of ML techniques, it has been widely recognized that they can lead to discrimination.
Many reasons can contribute to this, including biased sampling, incorrect labeling (especially with subjective labeling), biased representation (e.g., incomplete or correlated features), suboptimal or insensitive optimization algorithm, shift of population or data distribution, and failure to consider domain-specific, legal, or ethical constraints~\cite{Calders2013,hajian2016algorithmic}. 
Various techniques have been proposed to address these causes of ``unfair algorithms''~\cite{kamishima2012fairness,hajian2016algorithmic,zafar2017fairness,zemel2013learning,joseph2016fairness}.
For example Calders and \v{Z}liobait{\.{e}} suggested techniques to de-bias data~\cite{Calders2013}, including modifying labels of the training data, duplicating or deleting instances, adding synthetic instances, and transforming data into a new representation space. 

We use a recently proposed data de-biasing method that applies a preprocessor to transform the data~\cite{Calmon2017}. 
The result is a new dataset which is ``fairer''--while also limiting local deformations from the data transformation.
This is because the preprocessor optimizes data transformations with respect to penalties that rise with the magnitude of a feature change (e.g. changing a persons age from 5 to 60 will result in a higher penalty than from 5 to 8).
Simply put, if raw data contains biases that lead to an unfair model with a discriminatory feature (e.g., certain racial category is weighed more negatively than others), the data preprocessing mitigates the bias introduced by that feature.
This method has the benefit of retaining all features (as opposed to removing the discriminatory feature), which, among other benefits, would also allow exploration of correlations among them~\cite{Calders2013}.

The above debiasing techniques are \textit{normative} by nature, i.e., they rely on prescriptively defining the criteria of fairness in order to optimize for that criteria.
A recent paper pursued a complementary \emph{descriptive} approach by empirically studying how people judge the fairness of features used by a decision support system in the criminal justice system~\cite{Grgic-Hlaca2018}.
Their study uncovered the underlying dimensions in people's reasoning of algorithmic fairness, and demonstrated individuals' variations on these dimensions.

We adopt the same descriptive view, empirically studying how people judge fairness of an ML system and considering individual differences in their prior position on algorithmic fairness.
However, we also fill a gap in prior work by investigating how normative fairness (via the use of the preprocessor) is \textit{perceived} by people, and what factors impact such perception.

\subsection{Explanation of Machine Learning}
Explainable AI (XAI) is a field broadly concerned with making AI systems more transparent so people can confidently trust an AI system and accurately troubleshoot it --- fairness issues included.
Work on model explanations can be traced to early work on expert systems~\cite{swartout1985explaining,clancey1983epistemology}, which often explicitly revealed reasoning rules to end-users.
There has been a recent resurgence of XAI work driven by the challenge to interpret increasingly complex ML models, such as multi-layered neural networks, and by the evidence that ethical concerns and lack of trust hampers adoption of AI applications~\cite{Lee2018,hajian2016algorithmic}. 

A large volume of XAI work is on producing more interpretable models while maintaining high-level performance (e.g.~\cite{chen2016infogan,liang2017interpretable}), or on methods to automatically generate explanations. 
Given the complexity of current ML models, explanations are often \textit{pedagogical}~\cite{tickle1998truth}, meaning that they reveal information about how the model works without faithfully representing the algorithms. 
Many methods rely on some kind of sensitivity analysis to illustrate how a feature contributes to the model prediction~\cite{ribeiro2016should,lundberg2017unified}, so they can be model-agnostic, thus applicable to complex models.
For example, LIME explains feature contribution by what happens to the prediction when the feature-values change (perturbing data)~\cite{ribeiro2016should}.
Another common category is \textit{case-based explanations}, which use instances in the dataset to explain the behavior of ML models.
Examples include using counter-examples~\cite{wachter2017counterfactual} and similar prototypes from the training data~\cite{kim2016examples}.
Case-based explanations are considered easy to consume and effective in \textit{justifying} the decision, but may be insufficient to explain how the model works.

Work on how people perceive explanations of ML systems is a growing area~\cite{stumpf2007toward,lim2009and,abdul2018trends,Kulesza2013} which aims to inform the choices and design of explanations for particular systems or tasks.
Recent work calls for taxonomic organizations of explanations to enable design guidelines~\cite{lim2009and}.
In earlier work on explaining expert systems, researchers argued the difference between \textit{description v.s. justification}--by making not only the \textit{how} visible to users, but also the \textit{why}~\cite{swartout1985explaining}.
Accordingly, Wick and Thompson discussed the taxonomy of \textit{global-local} explanations~\cite{wick1992reconstructive}.
During initial practice, users may need global explanations that describe ``how the system works.''
During actual use, users tend to rely on justifications of why the system did what it did on particular cases.

Another useful taxonomy is proposed by Kulesza et al. by considering two dimensions of explanation fidelity: \textit{soundness} (how truthful each element in an explanation is with respect to the underlying system) and \textit{completeness} (the extent to which an explanation describes all of the underlying system)~\cite{Kulesza2013}.
They empirically showed that the best mental models arose from explanations with both high completeness and high soundness.
However, crafting highly complete explanations comes with a tradeoff, as completeness usually requires increasing the length and complexity of the explanation, which was shown to be detrimental to task performance and user experience in previous studies~\cite{Narayanan2018}. 

While researchers have explored user preferences in explanation styles, they have paid little attention to individual differences in such preferences.
Meanwhile, psychological research has long been interested in individual differences in explanatory reasoning. 
For example, research shows that some prefer simple, superficial explanations and others are more deliberative and reflective in their reasoning~\cite{fernbach2012explanation,klein2014influencing}.
Such individual differences can be predicted by cognitive style (e.g., cognitive reflection, need for cognition)~\cite{fernbach2012explanation} and culture~\cite{klein2014influencing}.
It is therefore possible that individuals differ in preferences for completeness and soundness of explanations.

Our work is concerned with how explanations impact fairness judgments of ML systems.
We build on a recent study by Binns et al., which examined human perception of a classifier's fairness in the insurance domain~\cite{Binns2018a}.
They provided four different explanation types applied to \textit{fictional scenarios} to elicit fairness judgment.
While the study provided rich qualitative insights on the heuristics people use to make fairness judgments, the authors acknowledge a lack of ecological validity as the explanations were not drawn from real ML model output.
Moreover, the explanations were not produced for the same data points, so they were incommensurate, which could possibly explain the absence of conclusive preference in their quantitative results.

Our work set out to overcome limitations on prior work by \emph{automatically generating} four types of explanations on a real ML model, and quantitatively examining how they impact people's fairness judgments.
Combining this advancement with the use of the data preprocessor allowed us to perform more carefully controlled experiments for ML fairness perception than prior work.

\section{Study Overview}

Related work informed four main considerations of our study: use case, choices of explanation styles, fairness issues we focus on, and the individual differences we explore.
Through both quantitative and qualitative results, we aim to answer the following RQs:
\begin{enumerate}[labelindent=15pt,labelwidth=\widthof{\ref{last-item}},label=\arabic*.,itemindent=1em,leftmargin=!]
\item[\textbf{RQ1}] How do different styles of explanation impact fairness judgment of a ML system?
\begin{enumerate}[labelindent=20pt, itemindent=1em,leftmargin=!]
\item[\textbf{RQ1a}] Are some explanations judged to be fairer? 
\item[\textbf{RQ1b}] Are some explanations more effective in surfacing unfairness in the model?
\item[\textbf{RQ1c}] Are some explanations more effective in surfacing fairness discrepancies in different cases?
\end{enumerate}

\item[\textbf{RQ2}] How do individual factors in cognitive style and prior position on algorithmic fairness impact the fairness judgment with regard to different explanations?

\item[\textbf{RQ3}] What are the benefits and drawbacks of different explanations in supporting fairness judgment of ML systems?
\label{last-item}\end{enumerate}

\subsection{Use Case: COMPAS recidivism data}
We conducted an empirical study with a ML model trained on a real data set.
Similar to~\cite{Grgic-Hlaca2018}, we chose a publicly available data set for predicting risk of recidivism (reoffending) with known racial bias\footnote{\url{https://www.kaggle.com/danofer/compass}}. The data set was collected in Broward County, Florida over a two year span. It is used by COMPAS (Correctional Offender Management Profiling for Alternative Sanctions), a commercial algorithm to help judges score criminal defendants' likelihood of reoffending.
However, ProPublica has reported on troubling issues with the COMPAS system~\cite{Larson2016, ProPublicaData}.
First, the classifier may have low overall accuracy (\cite{ProPublicaData} reported 63.6\%).
Second, the model is reported to exhibit racial discrimination, with African American defendants' risk frequently overestimated.

We chose a criminal justice use case because it carries weight to elicit reaction on fairness, even for the general population.
Note our goal is not to study the actual users of COMPAS. 
Rather, we are using the use case as a ``probe'' to empirically study fairness judgments.
The same use case was used in previous studies to understand how people perceive algorithmic fairness with regard to features used~\cite{Grgic-Hlaca2018}.

\subsection{Explanation Styles}
We chose to programmatically generate the four types of explanations introduced by Binns et al.~\cite{Binns2018a} (details to be discussed in System Overview section) because they represent a set of common approaches in recent XAI work.
They  embody the categorization of \textit{global v.s. local} explanations.
Specifically, influence and demographic-based explanations are global styles as they describe how the model works; sensitivity and case-based explanations are local styles as they attempt to justify the decision for a specific case.
These explanation styles vary along the taxonomy introduced by previous work in other ways: e.g. sensitivity based explanation is similar to a ``what if''~\cite{lim2009and} and the case-based explanation is the least ``sound'' of the explanation types discussed~\cite{Kulesza2013}.

\subsection{Fairness Issues}
Given our use case, we consider fairness issues in terms of racial discrimination.
While there are other controversial features in the dataset~\cite{Grgic-Hlaca2018}, race is generally considered inappropriate to use in predicting criminal risk (termed \textit{protected variable}).
We focus on two types of fairness issues.

\subsubsection{Model Unfairness} 
As discussed, the COMPAS data set is known to be racially biased, but we mitigate that bias by using the data processing method in~\cite{Calmon2017}.
In the experiment, we introduce the use of the \textbf{data processing} technique as a between-subject variable.
By comparing participants' fairness judgment for a model trained on the \emph{raw} data to that of \emph{processed} data, we aim to understand whether participants could identify the model-wide fairness issue, and whether certain explanations expose the problem better. 

\subsubsection{Case-specific disparate impact} Predictions from an ML algorithm are not uniformly fair--consider \textit{disparate impact} from a protected variable.
For example, if two individuals with identical profile features but different racial categories receive different predictions, it should be considered unfair~\cite{Calders2013, Grgic-Hlaca2018}.
Statistically, these cases are on the decision boundary of the feature space given the relatively small weight of the race factor, meaning they have low-confidence predictions that may be unfair.
In the experiment, we introduce \textbf{disparate impact} by race factor as a within-subject variable (i.e., each participant would be asked to judge some cases with disparate impact and some not). We adopt a factorial design with \emph{disparate impact} and \emph{data processing}.
For subjects given models after data processing~\cite{Calmon2017}, disparate impact is reduced.
We aim to discover how well participants identify the case-specific fairness issues using different explanations. 

Our hypothesis is that given local explanations focus on justifying a particular case, they should more effectively surface fairness discrepancies between cases.
In contrast, global explanations may require additional effort to reason about the position of the case with respect to the decision boundary (e.g., ``\emph{This person's features all have no impact in the model, except race}'').
Note that local explanations may expose the case-specific fairness issue differently.
Case-based explanation exposes the boundary position with a low percentage of cases justifying the decision. 
Sensitivity-based explanation explicitly describes disparate impact--``\emph{Changing this person's race changes the prediction}''.


\subsection{Individual Difference factors}
Based on prior work, we focus on two areas of individual factors: cognitive style and prior position on algorithmic fairness.
For cognitive style, we measure individual's \textit{need for cognition}~\cite{cacioppo1984efficient}.
For prior positions, we consider two levels: \textit{general position on the fairness of using ML systems} for decision support, and \textit{position on the fairness of using a particular feature}--here we focus on the race factor.

\section{System Overview}

\subsection{Re-offending Prediction Classifier}
The model is a binary classifier predicting whether an individual in the COMPAS data set is likely to re-offend or not, implemented by Scikit-learn's logistic regression.
The use of a regression model is ecologically valid---many current decision support systems use such simple and interpretable models~\cite{Veale2018}.
However, the explanation styles we study are not limited to regression models.

We built the model using a subset of features in the COMPAS dataset\footnote{We split the data into training set (4222 samples) and testing set (1056).},
including \textit{Race} as the feature with fairness issues.
For simplicity, we focus on two racial groups (Caucasian and African-American), and filtered others.
Other features included: \textit{Age} (18-29/30-39/40-49/50-59/>59), \textit{Charge Degree} (Felony/Misdemeanor), \textit{Number of Prior Convictions} (0/1-3/4-6/7-10/>10), and \textit{Had Juvenile Convictions} (True/False).
According to Grgic-Hlaca et al.~\cite{Grgic-Hlaca2018}, charge degrees and criminal history were deemed fair in a similar use case.
Age is also important in assessing re-offense risk.
Following statistical convention, all categorical features are dummy coded using the median category as the reference level, where possible.

The accuracy of the model is 67.1\% on raw data and 67.6\% on processed data, comparable to reports on the accuracy of COMPAS system~\cite{NorthPointeValid, Larson2016}.
Note that logistic regression also produces a confidence level implicitly in class probabilities.

\subsubsection{Data processing and cases with disparate impact}
We used the method introduced in~\cite{Calmon2017} to perform data processing then re-trained the model. 
The resulting model reduced bias against the African American group, as evidenced by the feature co-efficient being reduced from  0.177 to -0.036 (A feature co-efficient of 0 corresponds to the feature having no effect in the decision).

\begin{figure*}
	\centering
	\vspace{-5pt}
\includegraphics[width=.99\textwidth]{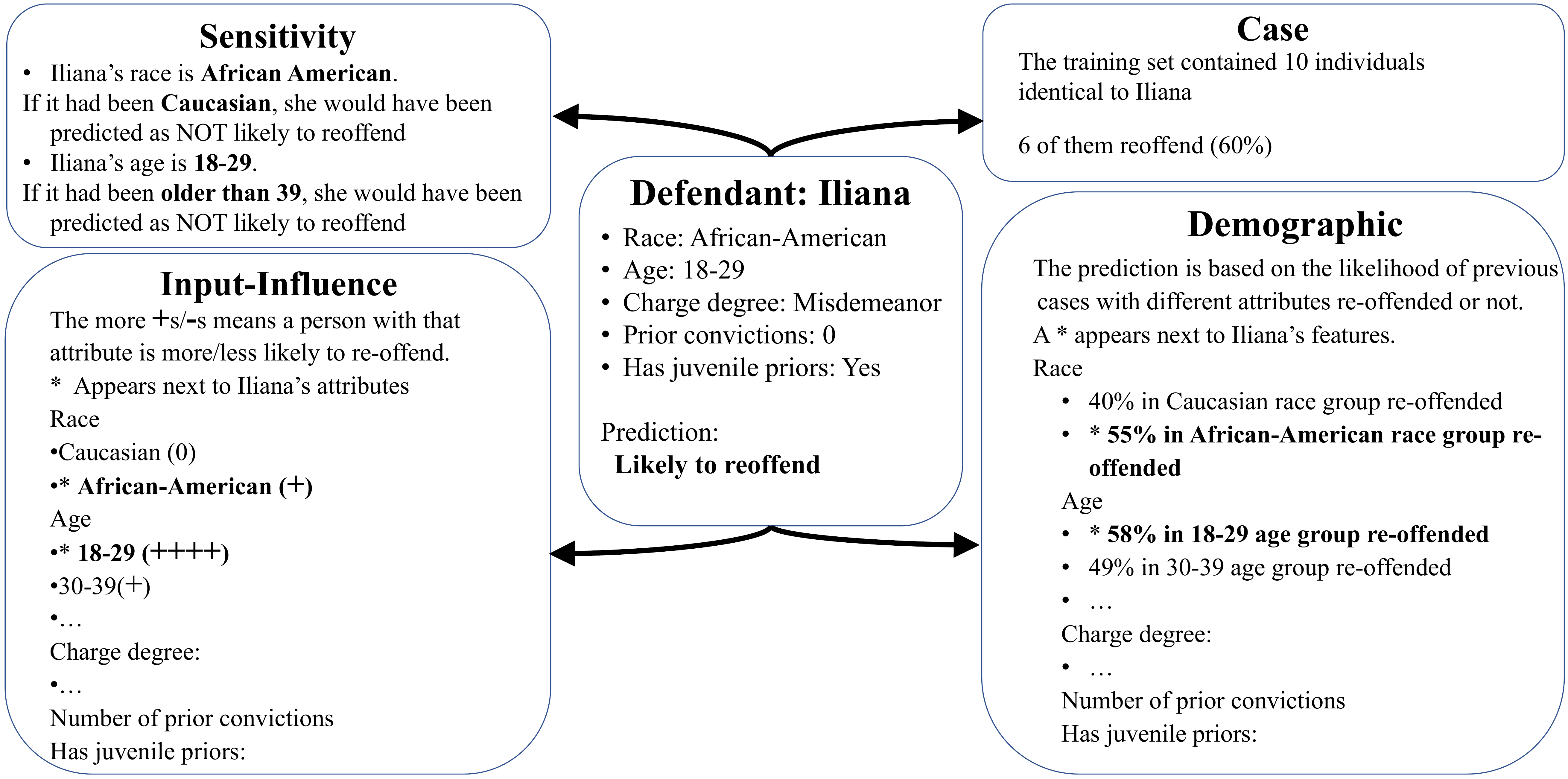}
\vspace{-10pt}
	\caption{Examples of explanations generated using the raw data classifier, adjusted and truncated for brevity.
    Consult our supplementary materials for full explanation output from both classifiers -- as seen by participants (including * and highlights).
    }
    \Description{Examples of explanations generated using the raw data classifier, adjusted and truncated for brevity.}
	\label{figExplanations}
    \vspace{-5pt} 
\end{figure*}

To identify cases with unfair treatment of disparate impact, we follow the definition  ``\textit{treating one person less favorably on a forbidden ground than another...in a comparable situation}''~\cite{Calders2013}.
That is, if perturbing a test example's protected variable (race) changes the algorithm's prediction, we consider it to have disparate impact.
We found 23 cases in the raw dataset--all very near to the decision boundary\footnote{The raw data classifier's confidence on the disparately impacted sample group had an average of 52\% and a max of 54\%.
The processed data classifier had average and max confidence both at 50\%.}. 

\subsubsection{Sampling cases for the user study}
Due to user study time constraints, we could only show each user a small sample of the explanations.
Since we intended to study fairness discrepancies between disparately impacted and non-impacted cases, we over-sampled the former category.
Among the 23 disparately impacted instances, we sampled all 8 unique cases (i.e. the rest had the same feature-values as one of the 8 we sampled).
From the non-impacted group of 992 instances, we sampled 16 unique cases.

\subsection{Explanation Generation}
As discussed, we patterned our explanations, shown in Figure~\ref{figExplanations} (truncated version, see supplementary materials for the full version), after the templates presented by Binns et al.~\cite{Binns2018a}.
While Binns et al. manually created examples of these explanation, we developed programs to automatically generate them to obtain comparable explanation versions for the same data point, controlling for differences in representation and presentation.
These generation methods can also be broadly applied to ML prediction models using relational features.




\subsubsection{Input Influence-based Explanation}

describes the decision boundary itself.
Because the feature coefficients of the logistic regression model encode the relative importance of each feature, we present them as strings of `+' and `-' in our explanations, as shown in Figure~\ref{figExplanations}.
To do this, we discretized them into 11 buckets, based on the range of the maximum and minimum coefficient.
This type of explanation is \emph{global} since the decision boundary is a property of the classifier, and thus is described the same way for all samples.

\subsubsection{Demographic-based Explanation}

describes the structure of the training data and how it is distributed with respect to the decision boundary.
We simply summarize, for training data matching each feature category, the percentage with the same label as predicted for the presented example.
This type of explanation is \emph{global} and generates the same description for all samples on each side of the decision boundary.

\subsubsection{Sensitivity-based Explanation}

seeks to modify the presented sample along each feature until the prediction changes.
When the prediction does change, we report back to the user the necessary feature change to produce the change in output.
This type of explanation is \emph{local} as it is specific to each presented example, and justifies the decision by indicating changes needed to produce a different output.

\subsubsection{Case-based Explanation}

does a nearest neighbor search in the training data to find similar cases.
Since our study has a large data set with respect to the feature space, we frequently find neighbors occupying the same feature space location as the sample presented for explanation.
When this is the case, we show the \% of those neighbors with the same label as the prediction.
When no exact matches are found, we simply show the features and label for the nearest neighbor in the training data.
This is a modification to the design in Binns et al., which  describes only a single identical or similar case.
This explanation is \emph{local}, and attempts to justify the decision by indicating similar examples with similar outputs.

\section{Methodology}
Our study adopted a mixed design by having data processing (raw or processed) and explanation styles (4 styles) as between-subject variables, and disparate impact as a within-subject variable. 
Each participant completed 6 fairness judgment trials in a random order, where each trial consisted of judging a single case. 
2 trials were randomly selected from the 8 disparately impacted cases, and 4 trials from the 16 non-impacted cases in the test data.

In September 2018 we recruited 160 Amazon Mechanical Turk workers, with the criteria that the worker must live in the US and have completed more than 1000 tasks with at least a 98\% approval rate.
They were randomly assigned to the 8 conditions (2 data processing treatments $\times$ 4 explanations).
Among them, 62.5\% are male, 78.8\% are self-identified as Caucasian, 29.4\% are under 30, and 13.3\% are above 50.
In our quantitative analysis, we included participants' race and gender as co-variance but observed no significant effect.

\subsection{Study Procedure}
We conducted an online survey-style study. 
Participants first provided informed consent, then reviewed a scenario about a ML system developed to assess defendants' re-offense risk, to help a judge make bail decisions.
Each participant was given 6 trials, and within each trial, the participant would first review information about an individual (Figure~\ref{figExplanations}, center region), then make his or her own prediction about whether that individual would re-offend.
This step was to ensure they carefully consider the features used in the prediction.
Subsequently, they would be presented with the prediction from the machine learning classifier with one of the four styles of explanations.
They were asked to rate agreement with: ``\emph{How the software made the prediction was fair}'' based on a 1 (Strongly Disagree) to 7 (Strongly Agree) Likert Scale, together with an open question justifying their rating.


We checked attention after the third trial, by asking participants to find the feature not used by the software from a list of choices.
All participants passed the check.
After completing all tasks, participants would answer a survey measuring their individual differences, and collecting demographic data.
On average the study took 18 min to complete, and each participant was compensated with \$3.

\subsection{Individual Differences}
We measured two types of individual differences: prior positions on algorithmic fairness, and cognitive style. 
For prior positions, we further differentiate between one's general position on using ML systems to assist decision-making, and position on the fairness of using race as a feature.
We use a semantic differential scale that is often used to measure attitude on controversial topics~\cite{liao2014expert}.
Specifically, we ask participants to rate the statements ``\emph{Considering race as a factor in criminal risk prediction}'' and ``\emph{Using machine learning software to replace or augment human decision making}'', on 5 pairs of bipolar adjectives such as ``\emph{harmful-beneficial}'' and ``\emph{unethical-ethical}'' with a 7-point Likert scale.
The ratings of the 5 pairs are averaged to represent the individual's prior position.
The Cronbach Alpha for the scale on general ML position is 0.95, and for the race feature is 0.98, showing excellent consistency.
For cognitive style, we chose to measure one's Need for Cognition using only 4 of the items from the standard scale~\cite{cacioppo1984efficient}, due to the time constraints. 
\section{Results: Quantitative}

We start with examining the effect of explanation style, data processing (\emph{raw/processed}) and disparate impact (\emph{true/false}) on participants' fairness judgment (RQ1).
We then explore how individual differences, including prior position on fairness of ML, prior position on fairness of using the race feature, and need for cognition, impacted the judgment (RQ2).
All statistical analyses were done in R.
The \texttt{lmerTest} package was used to run mixed-model regressions.

\subsection{Explanation, data processing, and disparate impact}
\begin{figure}[t]
    \centering
     \vspace{-5pt} 
     \includegraphics[width=\columnwidth]{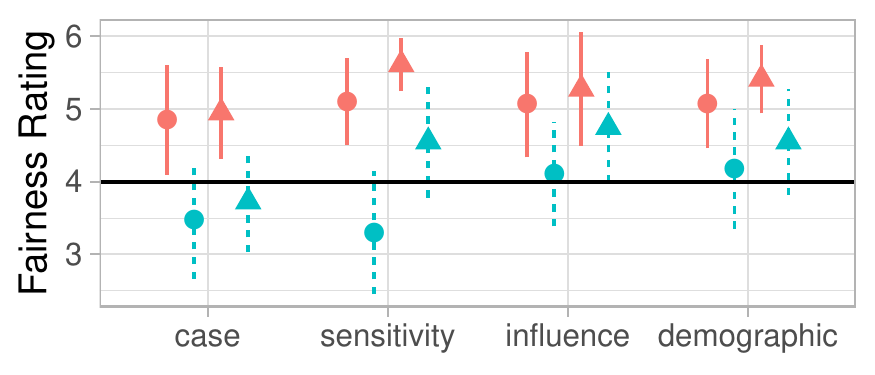}
     \vspace{-15pt}
    \caption{Overall mean ratings of fairness, per explanation type, data process treatment \emph{(raw=$\CIRCLE$, processed=$\blacktriangle$)}, and sample group \emph{(impacted=blue dashed lines, non-impacted=red solid lines)}.
    The lines indicate the 95\% confidence intervals.}
    \Description{Overall mean ratings of fairness, per explanation type, data process treatment, and sample group.}
    \label{fairness}

     \vspace{-10pt}
     
\end{figure}

Given the complexity of the statistical model, we first describe the trends with the descriptive data, then report statistical testing results. 
In Figure~\ref{fairness}, we plot the mean and the 95\% confidence interval of the mean of fairness ratings in all experiment treatments, showing several trends: 
\begin{enumerate}[labelindent=15pt,labelwidth=\widthof{\ref{last-item2}},itemindent=1em,leftmargin=!]
\item Predictions made on the processed data (triangles) were rated fairer than those on the raw data (circles).
It suggests that participants perceived fairness issues for the model trained on the raw data, and the processing technique mitigated the problem.
\item Predictions made on cases with disparate impact (blue dashed lines) were rated less fair than those without it (red solid lines). 
This shows participants' fairness perceptions align with the presence of a fairness discrepancy between groups.
\item Explanation styles made nuanced differences.
As expected, the two local explanations led to higher discrepancy of fairness ratings between disparately impacted cases and non-impacted cases (difference between the dashed and solid lines) than the two global explanations.
Thus, the former are more effective in exposing case-specific fairness issues. 
Moreover, this difference is most prominent for sensitivity-based explanations applied to raw data.
This could be caused by sensitivity-based explanation being the most explicit in exposing disparate impact, while data processing mitigated the problem.
\label{last-item2}\end{enumerate}

We now report the statistical significance of these observed trends. 
In particular, to validate that sensitivity-based explanation is most effective in exposing the disparate impact issue in the raw data, we expect to see a three-way interaction between explanation style, data processing, and disparate impact.  
We construct a mixed-effect regression model with the three-way interaction (and all the lower order interactions) as fixed effects, and participant as a random effect.
We control for gender and race as covariances and neither has significant effect.
The three-way interaction we expected is \textit{not} significant, $F (3, 152)=0.54$, $p=0.66$.
There is a marginally significant\footnote{We consider $p<0.05$ as significant, and $0.05 \leq p<0.10$ as marginally significant, following statistical convention~\cite{cramer2004sage}} two-way interaction between explanation style and disparate impact, $F (3, 152)=2.35$, $p=0.07$, and significant main effect of disparate impact, $F (1, 152)=103.25$, $p<0.001$, and data processing, $F (1, 152)=4.65$, $p=0.03$. 

The main effect of data processing and disparate impact prove statistical significance for the first two observed trends.
The two-way interaction indicates that explanation styles had differential impact on exposing the disparate impact issue.
We conduct pairwise comparison for this interactive effect to identify between which explanation styles this perceived fairness discrepancy significantly differ.
We found that if we use sensitivity-based explanation as the reference level, influence-based explanation is significantly different, $F (1, 156)=5.14$, $p=0.02$, and demographic-based explanation is marginally significant, $F (1, 156)=3.29$, $p=0.07$; If we use case-based explanation as the reference, influence-based explanation is marginally different, $F (1, 156)=3.36$, $p=0.07$.
This validates the observation that local explanations are more effective than global ones in exposing fairness discrepancies in different cases.

While we did not find statistical significance of the three-way interaction that validates the effectiveness of sensitivity-based explanation, a possibility is that there are individual differences for which the model did not account. 
In the next section, we explore that possibility.

\begin{figure}[t]
    \centering
     \vspace{-5pt} 
     \includegraphics[width=0.495\columnwidth]{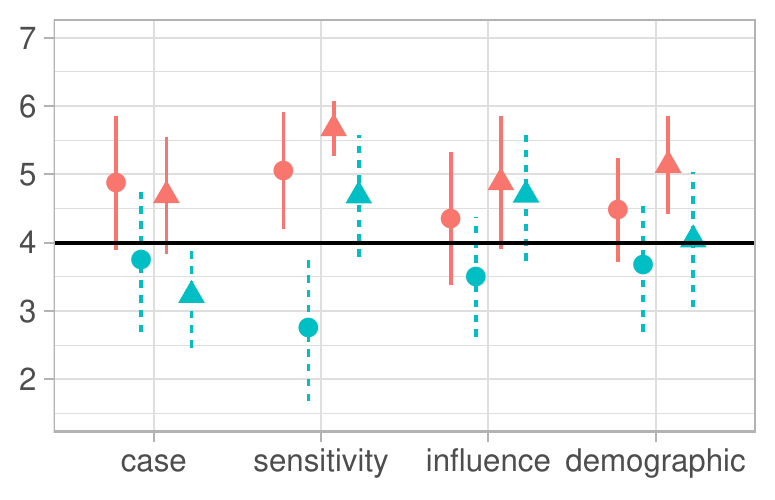}
     \includegraphics[width=0.495\columnwidth]{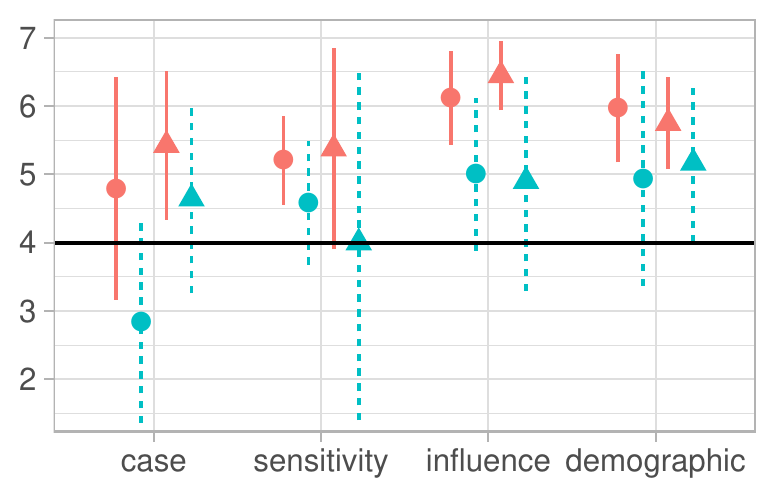}
     \vspace{-15pt}
    \caption{Same data as Figure~\ref{fairness}, split by prior position on the fairness of using the race feature.
    Left: Participants that consider using race ``Unfair'' \emph{(race\_pos < 4)}.
    Right: Participants that consider using race ``Fair'' or neutral \emph{(race\_pos >= 4)}.}
    \Description{Same data as Figure~\ref{fairness}, split by prior position on the fairness of using the race feature.}
    \label{fairnessRace}
     \vspace{-10pt}
     
\end{figure}

\subsection{Individual differences}
We enter the following factors into the model: prior position on using machine learning to assist decision-making (\textit{ML position}), prior position on fairness of using the race feature (\textit{race position}), and \textit{need for cognition}.
We start from four level interactions of each of the individual difference factors with the three manipulated variables (explanation, data processing, disparate impact), and then iteratively reduce it to lower-level interactions if it is not significant.
We eventually arrive in a model with the following terms: a four way interaction between race position and the three manipulated factors, $F (3, 144)=2.59$, $p=0.05$, and a marginally significant two-way interaction between ML position and explanation style, $F (3, 137)=2.43$, $p=0.07$.
We did not find need for cognition to make a difference and removed it.

By including these individual difference factors in the model, we now find the three-way interaction between explanation style, data processing, and disparate impact to be significant, $F (3, 144)=2.96$, $p=0.03$ (its lower-level two-way interactions as well).
In addition to the main effect of data processing ($F (1, 137)=4.68$, $p=0.03$) and disparate impact ($F (1, 144)=28.86$, $p<0.001$) as in the original model, we also find a main effect of ML position ($F (1, 137)=17.31$, $p<0.001$), race position ($F (1, 137)=6.43$, $p=0.01$), and a marginally significant main effect of explanation style, $F (3, 137)=2.11$, $p=0.10$. 

The above significant three-way and four-way terms, after including race position in the analysis,  demonstrate that the consideration of this individual factor ``de-noised'' the data.
In other words, it is only when an individual considers using race to be unfair, that a sensitivity-based explanation like this--``\textit{If Nolan had been `Caucasian', he would have been predicted to be NOT likely to re-offend}''--heightens the concern and significantly lowers the perceived fairness.
When an individual does not consider it problematic to use  race as a decision factor, they would not perceive such an explanation negatively.
This trend is illustrated in Figure~\ref{fairnessRace}, where we separate participants who considered the race factor unfair and those who considered it fair-to-neutral (33.1\% of all participants).
In fact, for those who consider race to be a fair or neutral feature to use (Figure~\ref{fairnessRace}, Right), they did not perceive predictions made on the raw data (circles) to be less fair than processed data (triangles), and generally rated fairness to be higher (thus the main effect of prior position on race).

\begin{figure}[t]
    \centering
     \vspace{-5pt} 
     \includegraphics[width=\columnwidth]{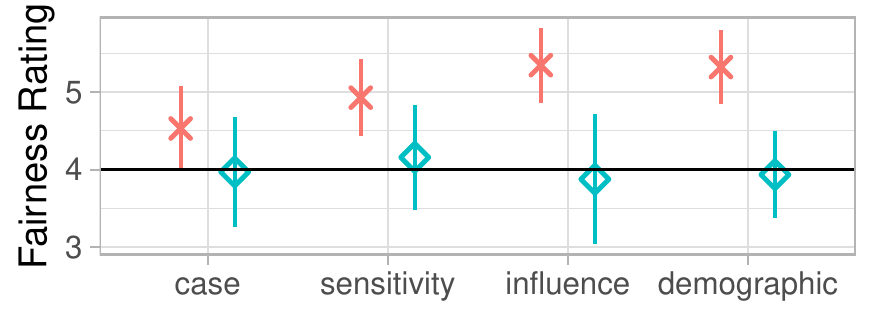}
     \vspace{-15pt}
    \caption{Overall mean fairness ratings, broken down by prior position on ``Trust in ML'' \emph{(high trust=$\times$, low trust=$\Diamond$)}.}
    \Description{Overall mean fairness ratings, broken down by prior position on ``Trust in ML''.}
    \label{trust}
     \vspace{-10pt}
\end{figure}

The main effect of ML position and its interactive effect with explanation style indicates that a general positive position on algorithmic fairness enhanced perceived fairness, and also led to different explanation preferences.
We conducted pairwise comparison between styles of explanation, and found this interactive effect with ML position to be significant for influence-based explanation, $F (1, 148)=6.25$, $p=0.01$, and marginally significant for demographic-based explanation, $F (1, 148)=2.77$, $p=0.10$, if using case-based explanation as the reference.
It is significant for influence-based explanation, $F (1, 148)=3.73$, $p=0.05$, if using case-based explanation as the reference.
This implies that people who trust ML systems gain \emph{even higher} confidence in the fairness of a prediction given global explanation (Figure~\ref{trust}).

It is worth noting that after controlling for these individual factors, we now see a marginally significant main effect of explanation style.
Pairwise comparisons show that case-based explanation was rated marginally significantly less fair than influence-based ($F (1, 153)=3.51$, $p=0.06$) and demographic-based explanation ($F (1, 148)=3.20$, $p=0.08$).
We consider it as evidence that case-based explanation is seen as generally less fair.

To summarize, in response to RQ1 and RQ2, we found evidence that: 
1) Case-based explanation is seen as generally less fair; Global explanations further enhance perceived fairness for those who have general trust for machine learning systems to make fair decisions. 
2) Local explanations are more effective than global explanations at exposing case-specific fairness issues, or fairness discrepancies between different cases.
Sensitivity-based explanations are the most effective in exposing the fairness issue of disparate impact made by a particular feature---but only if the individual views using that feature as unfair. 
3) In general, we show that individuals' prior position on ML trust and feature fairness have significant impact on how they react to explanations, and possibly more so than differences in cognitive styles.

\section{Results: Qualitative}
Along with collecting fairness ratings, we asked participants to justify their judgment. 
The authors reviewed this data and used open coding to extract themes in the answers.
Here we discuss two groups of themes.
One is to understand how participants made fairness judgments.
Another is on participants' feedback for the four styles of explanations. 

\subsection{How is fairness judgment made?}
In the open-ended answers, we investigated the criteria participants used to judge fairness.
We see variations in reliance on the provided explanations, and depth of reasoning about the algorithm's processes, providing further evidence of individual differences in the \textit{criteria} used to make fairness judgments of ML systems. 

\subsubsection{General trust or distrust in ML systems} 
Some participants provided reasons not specific to a case or explanation, but that general trust or distrust of ML systems dominated their judgment, as they tended to give consistent ratings across cases.
Reasons for a general trust include ``\textit{based on objective data is better than subjective opinions}'' (\user{31}{C}{R})\footnote{Participant IDs give treatment info, explanation (\textbf{\underline{S}}ensitivity, \textbf{\underline{C}}ase, \textbf{\underline{I}}nput-Influence, \textbf{\underline{D}}emographic) followed by data processing (\textbf{\underline{R}}aw, \textbf{\underline{P}}rocessed).}, ``\textit{large data set}'' (\user{37}{C}{R}), ``\textit{uses statistics based on prior knowledge to make a judgment}'' (\user{176}{I}{R}).
In contrast, some participants considered generalization by statistics unfair--``\textit{it might be unfair to group everybody together - makes more sense for the judge to have individual judgment.}'' (\user{184}{I}{P}), while others think that ``\textit{there needs to be a human element to the decision}'' (\user{62}{C}{R}).
These observations corroborate Binns et al.~\cite{Binns2018a} and further validate that participants varied on their general position on using ML system for criminal justice, and it influenced their fairness judgment.

\subsubsection{Features used} 
Participants frequently cited features used by the algorithm as reasons for fairness or unfairness. 
Some explicitly differentiated between the process of the algorithm and the feature considered--``\textit{The software makes it's decisions based on it's algorithm, so I believe it is fair and impartial on that account.
However, some of the categories it is programmed to consider, such as age and race, are unfair}'' (\user{71}{S}{P}).
It is interesting to note that we observe individual differences in the position on the fairness of race feature in the qualitative results as well. 
While many participants called out the problem of considering race, a few participants who saw the processed data commented that ``\textit{[if] race was not a predictor [it] may not accurately reflect the reality}'' (\user{68}{D}{P}).
There is also some controversy on using age and juvenile priors as features.
Participants' comments echo results from a previous study~\cite{Grgic-Hlaca2018} showing that people consider multiple dimensions (e.g., relevance, disparate outcome, volitionality) in their judgments about the fairness of features used in decision-making algorithms, and individuals weigh these dimensions differently.

\subsubsection{Lacking features} 
As observed in \cite{Binns2018a}, several participants criticized the limited features used in our simple model.
Some suggested to have more detailed information on current features, such as ``\textit{frequency of priors or the interval of time since the last prior in order to get a more accurate assessment of what one's prior record means}'' (\user{53}{D}{P}). 
Others are less optimistic about the possible sufficiency of features to ensure fairness--``\textit{software cannot fully take into account environmental factors that cause people to go down a bad path}'' (\user{76}{C}{R}).

\subsubsection{Prediction process} 
Many participants based their fairness judgment on their understanding of the algorithms' process.
Some, especially those presented with global explanations, closely examined explanation details, e.g. ``\textit{Software seems to be flawed in major areas... improper weighing of distant vs recent past, and a questionable choice of how to evaluate probabilities in each case}'' (\user{119}{D}{R}).
Some also considered failure to account for external factors, e.g. ``\textit{the number may be relatively accurate for the race and charge degree categories, but if the [past] laws were different they would probably be higher}'' (\user{107}{D}{R}). Moreover, multiple participants attributed their low fairness ratings  to insufficient understanding of process, or ``\textit{`how' the data is used}'' (\user{172}{D}{R}).

\subsubsection{Data issue} 
A few participants questioned the underlying data used. 
Almost all of them were in either the demographic- or case-based explanation conditions, as these two styles leverage information about distributions of similar cases to explain the decision.
For example, ``\textit{`Not re-offend' rate for African Americans is a little  low. I think the percentage may be higher in reality... data could have been biased}'' (\user{107}{D}{R}).

\subsection{Explanation styles}
Below we summarize codes that are prominent for each explanation style. 
These results could help us better understand the benefits and drawbacks of each explanation style, and inform future work on designs of ML explanations.

\subsubsection{Influence based} 
It is a global explanation that faithfully describes how each feature contributes to the algorithm's decision-making process. 
We observed that this explanation prompted comments on details of the process, such as the weights of different features, and the trends with regard to different categories of a feature, e.g.  
``\textit{it is fair because it doesn't discriminate by race, but rather on age and prior convictions... if someone exhibits a behavior pattern it is likely to will continue, and I think people who are young are more apt to take risks}'' (\user{208}{I}{P}). 
On the one hand, detailed description of the algorithm process adds to the confidence in participants' judgments, which may help explain its enhancement of fairness perception among those trusting ML algorithms. 
On the other hand, it exposes more information to scrutiny, and is thus subject to critiques from the heterogeneous standards of fairness.

\subsubsection{Demographic based} 
This is a type of global explanation that does not expose the \emph{process} of the algorithm, but justifies the decision with the data distributions.
Sometimes, the distributions were seen as convincing, e.g. ``\textit{The high percentage of people with more than 10 prior convictions who end up reoffending was staggering, and justifies the prediction}'' (\user{57}{D}{P}).
Other times, participants found its explanation of the process inadequate, as the percentages do not clearly connect to an outcome--``\textit{The percentages aren't high enough.
It could go either way}'' (\user{157}{D}{R}).
Sometimes it also directed participants attention to the potential biases of underlying data.

\subsubsection{Sensitivity based} 
The main benefit of sensitivity-based explanation seems to be its conciseness and explicitly directing attention to features relevant to the particular decision.
It appears to be convincing and easy to process when a decision is uncontroversial --``\textit{The rationale is so basic (no prior offenses) that it has to be fair}'' (\user{138}{S}{R}).
``\textit{It's taking into consideration everything that we would and puts it into an easy to read manner}'' (\user{220}{S}{P}).
Consistent with our quantitative results, for  disparately impacted cases where the race factor is explicitly mentioned, sensitivity-based explanation heightens the concern and was perceived most negatively--``\textit{It says that in the same situation, if the offender were African-American rather than Caucasian, they would have been likely to offend. This is racial profiling and inaccurate in my opinion.}'' (\user{124}{S}{R}).

\subsubsection{Case based} 
As we found in the quantitative results, case-based explanation was judged to be the least fair---and the qualitative results provided reasons.
First, some found it to provide little information about \textit{how} the algorithm arrives at a conclusion.
Second, the number of identical cases and the percentage of cases supporting the decision are often considered too small to justify the decision--``\textit{It was unfair for the defendant because she was compared to only 22 other identical individuals... not to mention that only a little over 50\% reoffended.}'' (\user{61}{C}{R}).
This observation is consistent with Binns et al.~\cite{Binns2018a}, however, our work is based on the actual output of a ML model trained on a real dataset -- allowing us to  empirically show a limitation of case-based explanation\footnote{We found that 16\% of the test data exhibited the failure mode of \emph{contradicting} the claim ($<50\%$ of individuals with identical features share label). 
Meanwhile \emph{insufficient justification} of the claim (between 45\% and 55\% label matches) was quite common, with 24\% of the test data.
The prevalence of these failure modes indicates inherent ``unsoundness.''}. 
Lastly, we found variations in individuals' positions on the fairness of the ``explained process'' (as opposed to the actual algorithm process) to make decisions based on identical cases. 
While some people consider it to be fair to ``\textit{compare the  actions of people with similar history and backgrounds}'' (\user{200}{C}{P}), others questioned the underlying rationale such as  ``\textit{is anyone really identical if more things considered}'' (\user{201}{C}{P}).

\section{Discussion}

\subsection{Supporting different needs of fairness judgment}
The most important take-away from our study is that there are multiple aspects and heterogeneous standards in making fairness judgments, beyond evaluating \emph{features}, as studied in previous work~\cite{Grgic-Hlaca2018}.
Our experiment highlights two types of fairness issues: unfair models (e.g., learned from biased data), and fairness discrepancy of different cases (e.g., in different regions of the feature space).
Our qualitative results further illustrate that algorithmic fairness is evaluated by various dimensions including data, features, process, statistical validity, as well as broader ethical and societal concerns. 

Our results highlight the need to provide different styles of explanation tailored for exposing different fairness issues. 
For example, we show that local explanations are more effective in exposing fairness discrepancies between different cases, while global explanations seem to render more confidence in understanding the model and generally enhance the fairness perception. 
Hybridizing the two techniques reveals a possible human-in-the-loop workflow; using global explanations to understand and evaluate the model, and local explanations to scrutinize individual cases. 

It is critical to note that different regions of feature space may have varied levels of fairness and different types of fairness issues.
This calls for development of fine-grained sampling methods and explanation designs to better support fairness judgment of ML systems.
To that end, we envision an active-learning paradigm for fairness improvement, where the system interactively queries the human for fairness judgment of its predictions, together with explanation options, then optimizes the algorithm based on feedback.

Our qualitative results suggest another useful categorization of explanation styles: \textit{process oriented v.s. data oriented explanation}. 
The case- and demographic-based explanations we studied leverage information on data distribution to justify its decision but reveal less on \textit{how} the decision was made. 
Influence- and sensitivity-based explanations link each feature to the decision. 
We observe a general preference for process-oriented (\textit{how}) explanations, although a focus on data has the potential benefit of directing attention to issues in the data and dilutes the ``blame'' on the algorithms.

\subsection{Individual differences and descriptive fairness}
Another contribution of our study is to empirically demonstrate how individuals' prior positions on algorithmic fairness impact their reaction to different explanations. 
We differentiate between a general position on algorithmic fairness, and position on fairness of a particular feature used. 

The difference between normative (prescriptively defining what is fair) versus descriptive fairness and its implication for algorithmic fairness has been discussed in previous work~\cite{Grgic-Hlaca2018}.
Empirically, we show that even though race is considered a protected variable, individual positions on its fairness still vary (close to one third of participants considered it neutral or fair to use).
This indicates a lack of agreement on the meaning of moral concepts, a result Binns et al.~\cite{Binns2018a} hinted at qualitatively.
In different contexts, an algorithm developer may have to choose between a normative or a descriptive position of fairness, and it is important to be aware of the variation of fairness position in the population.
For example, if a ML system takes a normative position and aims to eliminate pre-defined biases based on people's feedback, it may need to account for their prior positions to weigh the feedback differently. 
It may be arguable whether explanation should always attempt \textit{soundness} and \textit{completeness} for all individuals.
On the other hand, if a system aims to provide optimal decision support for individual needs, it would be useful to provide mechanisms for individuals to express their prior positions as direct input for the algorithm (similar to the idea of active-learning by tuning features~\cite{raghavan2006active,settles2011closing}).

\subsection{Limitations}
We performed our study with crowdworkers, rather than judges who would be the actual users of this type of tool.
Additionally, there are many styles and elements of explanations not studied here.
One example is \emph{confidence}, which we declined to present to participants because we could not control for it.
\section{Conclusion}
Our work provides empirical insights on how different styles of explanation impact people's fairness judgment of ML systems, particularly the differences  between a global explanation describing the model and a local explanation justifying a particular decision. 
We highlight that there is no one-size-fits-all solution for effective explanation--it depends on the kinds of fairness issues and user profiles. 
Providing hybrid explanations, allowing both overview of the model and scrutiny of individual cases, may be necessary for accurate fairness judgment. 
Furthermore, we show that individuals' prior positions on algorithmic fairness influence how they react to different explanation types. 
The results call for a personalized approach to explaining ML systems. 
However, specific to fairness, ML systems may need to take a normative or descriptive position in different contexts, which may differentially require corrective or adaptive actions considering individual differences in their fairness positions.

\begin{acks}
Thanks to Bhanukiran Vinzamuri for assistance with the data preprocess.
This work was supported in part by DARPA \#N66001-17-2-4030.
This research was sponsored in part by the U.S. Army Research Laboratory and the U.K. Ministry of Defence under Agreement Number W911NF-16-3-0001.
The views and conclusions contained in this document are those of the authors and should not be interpreted as representing the official policies, either expressed or implied, of DARPA, the U.S. Army Research Laboratory, the U.S. Government, the U.K. Ministry of Defence or the U.K. Government. 
The U.S. and U.K. Governments are authorized to reproduce and distribute reprints for Government purposes notwithstanding any copy-right notation hereon.
\end{acks}

\bibliographystyle{ACM-Reference-Format}
\bibliography{references}

\end{document}